# The channel capacity and information density of biochemical signaling cascades

*Tatsuaki Tsuruyama**

**Department of Diagnostic Pathology, Center for Anatomical, Forensic Medical and Pathological Research, Kyoto University Graduate School of Medicine, Yoshida-Konoe-cho, Sakyo-ku, Kyoto 606-8315, Japan*

**Send correspondence to:** T. Tsuruyama
Department of Diagnostic Pathology, Kyoto University Graduate School of Medicine
1 Yoshida-Konoe-cho, Sakyo-ku, Kyoto 606-8501, Japan
Phone: +81-75-753-4473; Fax: +81-75-761-9591
E-mail: tsuruyam@kuhp.kyoto-u.ac.jp

**Abstract**

Living cell signaling systems include multistep biochemical signaling reaction cascades (BSCs) comprising modifications of molecular signaling proteins. Substantial data on BSCs have been accumulated in the field of molecular biology and the analysis of signaling systems requires qualitative evaluation. However, quantification of the information and channel capacity of BSCs has not been focused on from the perspective of information theory. In the current study, we aimed to derive basic equations for describing the channel capacity and information density of BSCs using the fluctuation theorem. From the results, channel capacity and information density can be described using the average entropy production rate when the signaling system is minimally redundant. The channel capacity could actually be calculated for the mitogen-activated protein kinase BSC when it was minimally redundant. This quantitative method of examination is applicable to the quantitative analysis of BSCs.

## I. INTRODUCTION

Biological homeostatic systems are characterized by cellular biochemical signaling reaction cascades (BSCs). Qualitative or semi-quantitative analysis of BSCs is a key approach to studying molecular and cellular biology. On the other hand, an analytical method based on the concept of a chemical reaction network recently produced important results using numerical simulation, and computational/automated methodologies have been developed to implement this concept [1-6]. However, the application of this for BSCs and biological events has been limited. One of the main reasons for the difficulty in the quantitative analysis of BSCs is the absence of a theoretical basis. In the current study, we aimed to illustrate a general scheme for the quantitative analysis of BSCs.

Herein, we use a simple model of BSCs as follows:

$$A_0 + A_1(R) + M \leftrightarrow S_1$$
$$S_1 + A_2 + M \leftrightarrow S_1 + S_2$$
$$\ldots$$
$$S_j + A_{j+1} + M \leftrightarrow S_j + S_{j+1} \qquad (1)$$
$$\ldots$$
$$S_{n-1} + A_n + M \leftrightarrow S_{n-1} + S_n$$
$$S_n + DNA \rightarrow S_n - DNA + RNA$$

In scheme (1), we consider an open homogeneous reactor in contact with chemostats of a signal mediator, *M*, which drive the BSC out of equilibrium.

An increase in the ligand $A_0$ triggers the BSC when it binds to the receptor (*R*) $A_1$, activates $A_1$ to $S_1$, and also induces the binding of *M*. Subsequently, $S_1$ interacts with and activates $A_2$ to $S_2$, which, in turn, binds *M*. More generally, each activated signaling molecule $S_j$ potentially activates $A_{j+1}$. Signaling terminates when $S_n$ translocates into the nucleus, binds to a specific region of genomic DNA, and promotes the subsequent transcription of RNA (Fig. 1).

## II. METHODS

We have formulated a BSC as a system in which the concentration of modified proteins fluctuates minimally around a steady-state value in the absence of a specific signal event. Recently, fluctuation analysis has greatly advanced through the discovery of the fluctuation theorem (FT) and Jarzynski's equality [3,7,8]. We aimed to apply these advances to the qualitative analysis of BSCs.

Once the signal event is triggered, a modified fluctuation will be amplified beyond the minimal fluctuation. The amplified modified fluctuation steps are assigned a step number *j* (1≤ *j* ≤ *n*) in which $S_j$ molecules participate in the *j*-th (1≤ *j* ≤ *n*) step of the BSC thus:

$$\mathbf{S_j}: S_j = S_j^e \rightarrow S_j^e + dS_j \qquad (2)$$

where *e* signifies the concentration of $S_j$ at the initial state. In this step, we set the mean entropy production rate to be $<\sigma_j>$. For instance, we can write a BSC thus:

**S₁ S₂ S₃ ..., S_{n-1} S_n** (2')

Here, $\mathbf{S_j}$ is the *j-th* signaling step having a certain duration $t_j$. The total duration of signal

events is:

$$T \triangleq \sum_{j=1}^{n} S_j t_j \equiv \sum S_j t_j \qquad (3)$$

We introduced the total number of signaling molecules $\Lambda$:

$$\Lambda = \sum S_j \qquad (4)$$

All of the signaling events $\Phi$ of total duration $T$ are to be considered *a priori* as equally probable. We obtain information $I$ for the all of the given signal events, which is derived from the above serial alphabetic sequences,

$$I = \log \Phi \qquad (5)$$

The rate at which information is transmitted in the channel is $I/T$. This limit defines the capacity of the channel:

$$C = \lim_{T \to \infty} \frac{1}{T} \log \Phi \qquad (6)$$

Here, we define the relative proportions $p_j$ using the total number of signaling molecules $\Lambda$:

$$p_j \triangleq \frac{S_j}{\Lambda}, \quad \sum p_j = 1 \qquad (7)$$

The logarithm of the total signal event number $\Phi$ is given using Stirling's approximation [9,10]:

$$\log \Phi = \log \frac{\Lambda!}{\prod_{j=1}^{n} S_j!} \simeq \Lambda(\log \Lambda - 1) - \sum_{j=1}^{n} S_j(\log S_j - 1) = -\Lambda \sum p_j \log p_j \qquad (8)$$

We can rewrite (3) using (7) in the main text so that

$$T = \Lambda \sum p_j t_j \qquad (9)$$

We collect equations (7), (8), and (9), and maximize $\Phi$ using two parameters, α and β, as

$$d \log \Phi - \alpha d \sum p_j - \beta dT = 0 \quad (10)$$

This is an application of Lagrange's nondetermined coefficient determination method. Differentiating (10), we obtain

$$d \log \Phi = -d\Lambda \sum p_j \log p_j - \Lambda \sum p_j (1 + \log p_j) dp_j \quad (11)$$

Substituting (10) into (11), we obtain

$$-d\Lambda\left[\sum p_j \log p_j + \beta \sum p_j t_j\right] + \sum dp_j\left[-\alpha - \beta\Lambda t_j - \Lambda(1+\log p_j)\right] = 0 \quad (12)$$

Because $d\Lambda$ and $dp_j$ are independent variables, we can write

$$\sum p_j \log p_j + \beta \sum p_j t_j = 0 \quad (13)$$

$$-\alpha - \beta\Lambda t_j - \Lambda(1+\log p_j) = 0 \quad (14)$$

Substituting (14) into (13) yields

$$\sum p_j(-1-\alpha/\Lambda) = 0 \quad (15)$$

To satisfy (15), we set the left hand side to

$$-1-\alpha/\Lambda = 0 \quad (16)$$

From (16) and (14), we obtain

$$-\log p_j = \beta t_j \quad (17)$$

$$\sum \exp(-\beta t_j) = 1 \quad (18)$$

Subsequently, let us consider the inverse signaling cascade. Here we define as follows:
$\mathbf{S_{-j}}: S_{-j} = S_j^e \to S_j^e - dS_j \quad (19)$

In this step, we set the mean entropy production rate to be $<\sigma_{-j}>$. The negative suffix $-j$ implies inverse signal transmission against the assumed polarity of the cascade. Taken together with the $\mathbf{S_j}$ signaling, a BSC is commonly described as comprising the following steps: the first events trigger the modified signaling $\mathbf{S_j}$ and, after a sufficiently long duration of demodification proceeds, the fluctuation decreases to a minimal value, and the system recovers to its initial status (Figure 2).

For the inverse cascade, using a parameter β', we have:

$$-\log p_{-j} = \beta' t_{-j} \quad (20)$$

Using the mean entropy production rate $<\sigma_{-j}>$ for a sufficiently long demodification duration (e.g., the reverse signaling duration) $t_{-j}$, the FT combines the BSCs $\mathbf{S_j}$ and $\mathbf{S_{-j}}$ to give:

$$\lim_{t_{-j}\to\infty} \frac{1}{t_{-j}} \log\left(\frac{p_j}{p_{-j}}\right) = \langle\sigma_{-j}\rangle \quad (21)$$

Above, we noted that $t_j << t_{-j}$ (20) gives:

$$p_j = p_{-j} \exp\langle\sigma_{-j}\rangle t_{-j} \quad (22)$$

Eqs. (17), (20), and (21) give:

$$\lim_{t_{-j}\to\infty} \frac{1}{t_{-j}}\left(\beta' t_{-j} - \beta_+ t_j\right) = \langle\sigma_{-j}\rangle \quad (23)$$

In these actual BSCs, $t_j \ll t_{-j}$, (23) gives:

$$\beta' \simeq \langle\sigma_{-j}\rangle \triangleq \sigma \quad (24)$$

Eq. (24) implies that the mean entropy production rates are independent of $j$, when the given BSC is minimally redundant. Therefore, we can rewrite (22) as follows:

$$-\ln p_{-j} = \sigma t_{-j} \quad (1 \leq j \leq n) \quad (22')$$

Using (23) and (24):

$$\beta = \frac{t_{-j}}{t_j}(\beta_- - \sigma) \simeq \frac{t_{-j}}{t_j}\varepsilon \quad (25)$$

where ε is a small number.

### III. Mutual Information in BSCs

Let us consider the mutual information in BSCs. In an actual BSC as depicted in Fig. 3, activation of the $j+1$-th signal is delayed following activation of the $j$-th step: when modification of $S_j$ occurs, modification of $S_{j+1}$ is not observed immediately. This is because a signaling molecule protein slowly diffuses in the cytoplasm and it takes a long time for it to interact with other signaling molecules. To an observer, modification of $S_j$ comes with the probability $ø_j$ that $S_{j+1}$ is unmodified, and the probability $(1-ø_j)$ that $S_{j+1}$ is modified. Here we set $\zeta_j = [ø_j \log ø_j + (1-ø_j) \log (1-ø_j)]$ and $\zeta_{-j} = [ø_{-j} \log ø_{-j} + (1-ø_{-j}) \log (1-ø_{-j})]$; $0 \geq \zeta_j \geq -1$, $0 \geq \zeta_{-j} \geq -1$. The entropy $H_j$ and the conditional entropies $H_j(S_j; S_{j+1})$ are given by:

$$H_j = -p_j \log p_j - p_{-j} \log p_{-j} \quad (26)$$
$$H_j(S_j; S_{j+1}) = \zeta_j p_j \quad (27)$$

The active molecule $S_j$ has the probability $ø_j$ of coming through unmodified and a $(1-ø_j)$ probability of being deactivated (Fig. 3). To choose $p_j$ and $p_{-j}$ in such a way as to maximize them, we introduce the following function $U_j$ using the undetermined parameter $\lambda$:

$$U_j = -p_j \log p_j - p_{-j} \log p_{-j} - \zeta_j p_j + \lambda(p_j + p_{-j}) \quad (28)$$

Then

$$\frac{\partial}{\partial p_j} U_j = -\log p_j - 1 - \zeta_j + \lambda \quad (29)$$

$$\frac{\partial}{\partial p_{-j}} U_j = -\log p_{-j} - 1 + \lambda \quad (30)$$

Setting the right-hand parts of (29) and (30) to zero and eliminating λ:

$$\log p_j + \zeta_j = \log p_{-j} \quad (31)$$

$$\therefore p_j = p_{-j} \exp(-\zeta_j) \quad (32)$$

Similarly, the inverse of reaction -j-th step (19), also depicted in Fig. 3, can be examined in a similar manner:

$$H_{-j} = -p_j \log p_j - p_{-j} \log p_{-j} = H_j \quad (33)$$

$$H_{-j}(S_j; S_{j-1}) = \zeta_{-j} p_{-j} \quad (34)$$

We have introduced the following function $U_{-j}$ with the undetermined parameter λ':

$$U_{-j} = -p_j \log p_j - p_{-j} \log p_{-j} - \zeta_{-j} p_{-j} + \lambda'(p_j + p_{-j}) \quad (35)$$

Then,

$$\frac{\partial}{\partial p_j} U_{-j} = -\log p_j - 1 + \lambda' \quad (36)$$

$$\frac{\partial}{\partial p_{-j}} U_{-j} = -\log p_{-j} - 1 - \zeta_{-j} + \lambda' \quad (37)$$

Setting the right-hands of Eq. (36) and (37) equal to zero and eliminating λ',

$$\log p_j = \log p_{-j} + \zeta_{-j} \quad (38)$$

$$\therefore p_j = p_{-j} \exp(\zeta_{-j}) \quad (39)$$

Comparing (21) and (39),

$$\zeta_{-j} = -\zeta_j \quad (40)$$

and we introduce mutual entropies:

$$\Delta H_j \triangleq H_j - H_j(S_j; S_{j+1}) \quad (41)$$

$$\Delta H_{-j} \triangleq H_{-j} - H_{-j}(S_j; S_{j-1}) \quad (42)$$

Considering the forward and inverse orientation of a signaling cascade, the net mutual entropy is given:

$$\Delta H_j - \Delta H_{-j} = \left(H_j - H_j(S_j; S_{j-1})\right) - \left(H_{-j} - H_{-j}(S_{j-1}; S_j)\right)$$
$$= -\zeta_j p_j + \zeta_{-j} p_{-j} = -\zeta_j p_j - \zeta_j p_{-j} \quad (43)$$

A comparison of (21), (34), and (22') gives:

$$\zeta_j = -\sigma t_j \quad (44)$$

Using (43) and (44), the mutual information for a pair of $j$-th and $-j$-th steps is:

$$\lim_{t_{-j} \to \infty} \frac{\Delta H_j - \Delta H_{-j}}{p_{-j} t_{-j} + p_j t_j} = \lim_{t_{-j} \to \infty} \frac{-\zeta_j p_j - \zeta_j p_{-j}}{p_{-j} t_{-j} + p_j t_j} = \lim_{t_{-j} \to \infty} \frac{\sigma t_j p_j (1 + \exp(-\sigma t_{-j}))}{p_j \exp(-\sigma t_{-j}) t_{-j} + p_j t_j}$$
$$= \sigma \quad (45)$$

In conclusion, the mutual information is given by the mean production rate of the inverse step, when the BSC system is redundant.

**IV. The channel capacity and information density of MAPK BSC**

BSCs have been studied extensively using models of MAPK pathways, in which the epidermal growth factor receptor, c-Raf, MAP kinase–extracellular signal-regulated kinase kinase (MEK), and kinase–extracellular signal-regulated kinase (ERK) are phosphorylated constitutively following treatment with cytokines and other reagents. This Ras-c-Raf- ERK BSC (RRE) and Ras-phosphatidylinositol-3 kinase-AKT BSC (RKT) are ubiquitous signaling pathways that convey cell mitogenic and differentiation signals from the cell membrane to the nucleus [11-22].

We aimed to evaluate redundancy in BSCs by adaptation into an actual signaling cascade for calculating the channel capacity using previously reported data. We obtained the following equation from Eqs (22) and (24) in reference to Figure 2:

$$\sigma t_{-j} = -\log p_{-j} \cong -\log \frac{\int_{t_j}^T dS_{-j}}{\int_0^T \sum_{j=1}^n dS_j} \triangleq -\log \frac{S_{-j}}{S} \quad (1 \leq j \leq 3) \quad (46)$$

The integration in the numerator in the third formula in Eq (46) was obtained from the area ratio of the integration of the inactivated molecule's concentration $dS_{-j}$ from $t_j$ to $T$ over the sum of integrations. As an example, we analyzed the data from

Petropavlovskaia *et al.* regarding the activation of BSCs in RIN-m5F rat islet cells stimulated by two types of ligand: full-length recombinant islet neogenesis-associated protein (rINGAP) and a 15-amino-acid fragment of INGAP-P [16]. The RRE and RKT BSCs were analyzed. In detail,

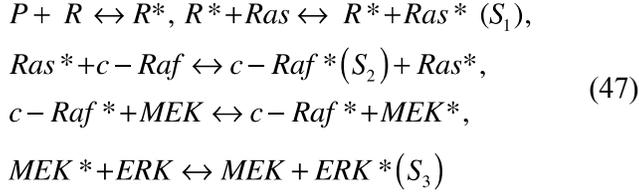

$$
\begin{aligned}
&P + R \leftrightarrow R^*, \ R^* + Ras \leftrightarrow R^* + Ras^* \ (S_1), \\
&Ras^* + c-Raf \leftrightarrow c-Raf^*(S_2) + Ras^*, \\
&c-Raf^* + MEK \leftrightarrow c-Raf^* + MEK^*, \\
&MEK^* + ERK \leftrightarrow MEK + ERK^*(S_3)
\end{aligned}
\quad (47)
$$

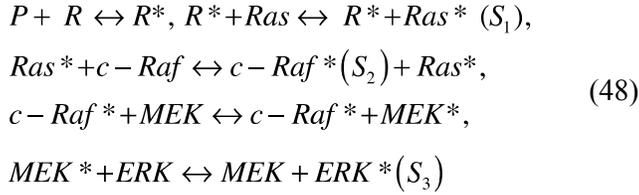

$$
\begin{aligned}
&P + R \leftrightarrow R^*, \ R^* + Ras \leftrightarrow R^* + Ras^* \ (S_1), \\
&Ras^* + c-Raf \leftrightarrow c-Raf^*(S_2) + Ras^*, \\
&c-Raf^* + MEK \leftrightarrow c-Raf^* + MEK^*, \\
&MEK^* + ERK \leftrightarrow MEK + ERK^*(S_3)
\end{aligned}
\quad (48)
$$

Here, * represents activated signaling molecules. $P$ and $R$ represent the peptide ligands and their receptor, respectively. $M$ represents a signal mediator. We plotted $-\log S_{-j}$ for BSC durations $t_{-j}$ ($j = 1, 2,$ and 3 for Ras, c-Raf, ERK), and it was possible to perform regression analyses for calculating the channel capacity $\sigma$ in Eq (24) from the gradient of the regression line. As a result, the RREs induced by INGAP-P and rINGAP were verified to function in a minimally redundant fashion ($n = 9$, $P = 2.3 \times 10^{-4}$, Fig. 4A; and $n = 6$, $P = 1.3 \times 10^{-2}$, Fig. 4B) and the channel capacities were calculated as $3.3 \times 10^{-2}$ bits/min and $1.7 \times 10^{-2}$ bits/min, respectively. In contrast, the RKTs induced by INGAP-P and rINGAP were not minimally redundant ($P > 0.05$, Fig. 4C; and $P > 0.05$, Fig. 4D) and the channel capacities were not computable.

## V. DISCUSSION

In this analysis, the equations for determining the basic properties of a nonredundant BSC, including information density and channel capacity, were newly formulated as Eq (45).

BSCs comprise complex networks and the reaction to such a network activates both the redundant and the minimally redundant BSCs. Our quantification methodology is available for distinguishing a given BSC from redundancy and could contribute to evaluation of the activity of a BSC in the whole cellular signaling network.

In conclusion, this framework provides a useful platform for quantitatively analyzing data on cell signaling pathways that are now being accumulated for quantitative modeling of biological systems.

## VI. METHODS

JMP Start Statistics version 9.0 (Statistical Discovery Software SAS Institute, Cary, NC, USA) was used for all analyses.

## ACKNOWLEDGMENTS

This work was supported by a Grant-in-Aid from the Ministry of Education, Culture, Sports, Science, and Technology of Japan (*Synergy of Fluctuation and Structure: Quest for Universal Laws in Non-Equilibrium Systems,* P2013-201 Grant-in-Aid for Scientific Research on Innovative Areas, MEXT, Japan). There are no conflicts of interest.

## FIGURE LEGENDS

**Figure 1. Schematic of a reaction cascade.** $A_0$ is a ligand and $A_1$ is a receptor that mediates cellular responses to external environmental changes. Individual signaling molecules $S_j$ (activated forms of $A_j$) relay the activation of individual steps and the last species $S_n$ is translocated to the nucleus, where it controls gene expression.

**Figure 2 A common time course of the *j*-th step.** Fold changes in modification. The ratio denotes the ratio at the steady state, normalized to 1. The integrals of $S_j = \int_0^{t_j} dS_j$ and $S_{-j} = \int_{t_j}^{T} dS_{-j}$ are shown, which satisfy $t_j << t_{-j} = T - t_j$

**Figure 3. Schematic representation of the relationship between inputs and outputs at the *j*-th signaling step of a simple discrete channel.**

**Figure 4. Quantification of the MAPK signaling pathway.**
(A, B) A regression analysis of log $S_j$ was performed for signal duration ($t_j$) (min) in the RRE with three steps. Regression lines are illustrated in the plots. The gradient of the regression line represents the channel capacity.
(C, D) A regression analysis was performed in the RKT.

**Figure 1**

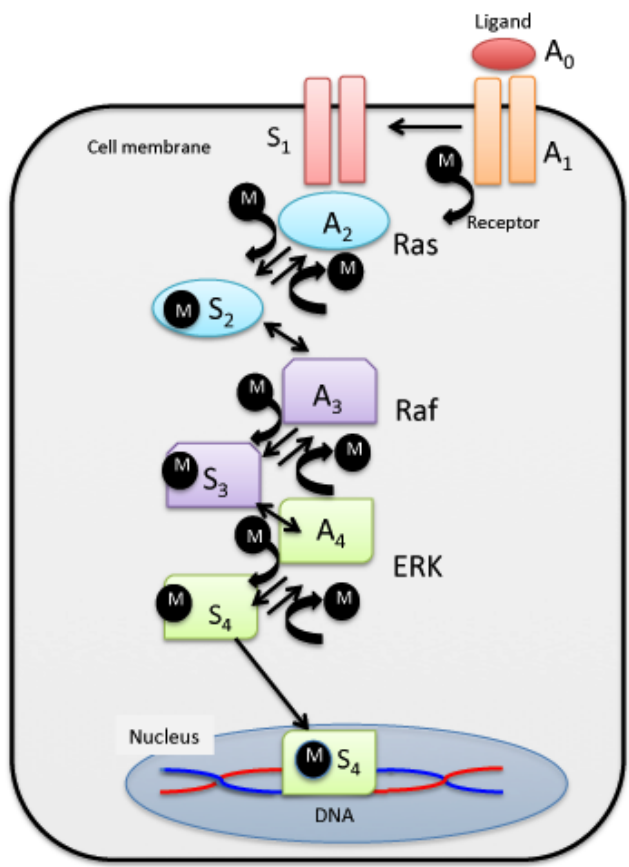

**Figure 2**

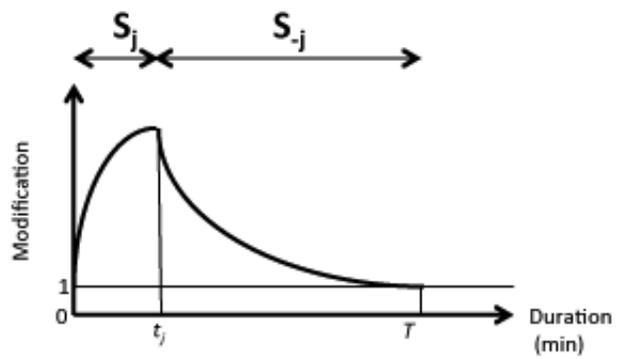

**Figure 3**

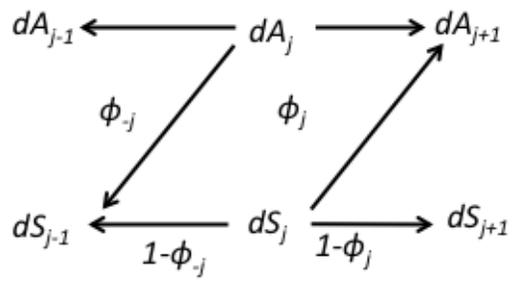

$\zeta_j = [\phi_j \log \phi_j + (1- \phi_j) \log(1- \phi_j)]$

$\zeta_{-j} = [\phi_{-j} \log \phi_{-j} + (1- \phi_{-j}) \log(1- \phi_{-j})]$

**Figure 4**

**A**

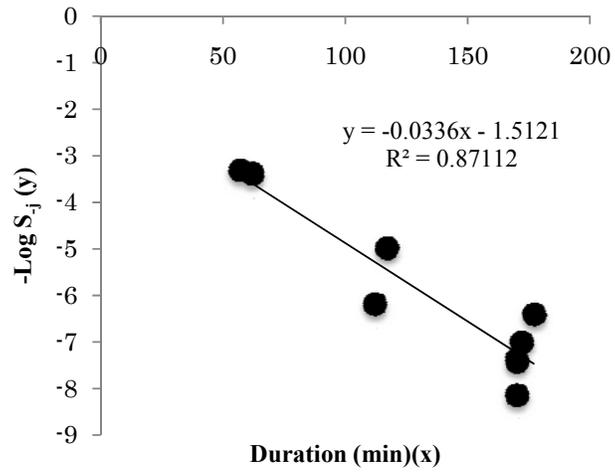

**B**

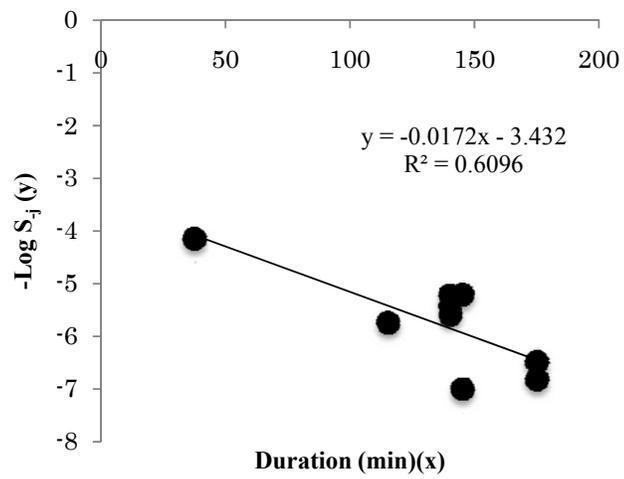

**C**

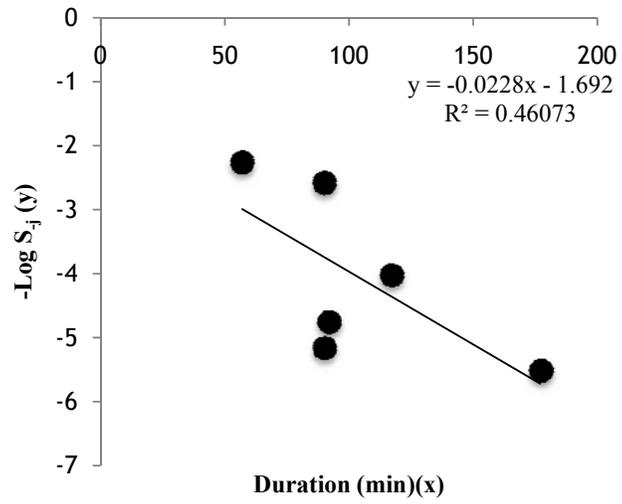

**D**

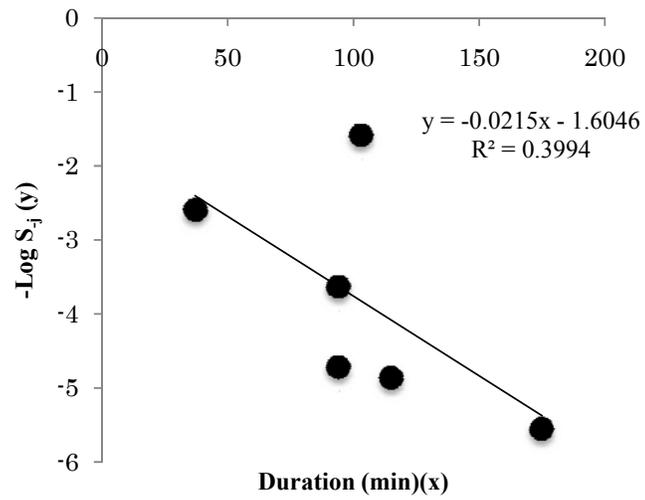